\documentclass[lettersize,journal]{IEEEtran}
\usepackage{amsmath,amsfonts}
\usepackage{algorithmic}
\usepackage{algorithm}
\usepackage{array}
\usepackage[caption=false,font=normalsize,labelfont=sf,textfont=sf]{subfig}
\usepackage{textcomp}
\usepackage{stfloats}
\usepackage{url}
\usepackage{verbatim}
\usepackage{graphicx}
\usepackage{cite}
\usepackage{listings}
\usepackage{subfig}
\usepackage{textcomp}
\usepackage{comment}
\usepackage{xcolor, soul}
\usepackage{multicol}
\usepackage{graphicx,epstopdf}
\usepackage{rotating}
\usepackage{array}
\usepackage{ragged2e}
\usepackage{cmap}
\usepackage{here}
\usepackage{float}
\usepackage{mathrsfs}
\usepackage[T1]{fontenc}
\usepackage{colortbl}
\usepackage{amssymb}
\usepackage{pifont}
\usepackage{epstopdf}
\usepackage[english]{babel}
\usepackage{amsmath}
\usepackage{epstopdf}
\usepackage{amsmath,epsfig}
\usepackage{lipsum}
\usepackage{subfig}
\usepackage{mwe}
\usepackage[justification=centering]{caption}
\usepackage{amssymb}

\usepackage{balance}

\usepackage[export]{adjustbox} 
\usepackage{cite}
\usepackage{amsmath,amssymb,amsfonts}
\usepackage{algorithmic}
\usepackage{graphicx}
\usepackage{textcomp}
\begin{document}


\title{Hybrid Machine Learning Models for Intrusion Detection in IoT: Leveraging a Real-World IoT Dataset}


\author{Md Ahnaf Akif, Ismail Butun, 
Andre Williams, and Imadeldin Mahgoub \\
\thanks{Corresponding authors: Ismail Butun (e-mail: ibutun@fau.edu), and Imadeldin Mahgoub (e-mail: mahgoubi@fau.edu).}
Department of Electrical Engineering and Computer Science,
Florida Atlantic University, Boca Raton, FL, USA \\
(E-mails: \textit{makif2022@fau.edu}, \textit{ibutun@fau.edu}, \textit{andrewilliam2018@fau.edu}, 
\textit{mahgoubi@fau.edu})

\thanks{This research is performed at the Tecore Networks Laboratory of Florida Atlantic University under the EECS Department and is funded by the Office of the Secretary of Defense (OSD) with Grant Number W911NF2010300.}
}

%


\IEEEpubid{0000--0000/00\$00.00~\copyright~2025 IEEE}

\maketitle

\begin{abstract}
The rapid growth of the Internet of Things (IoT) has revolutionized industries, enabling unprecedented connectivity and functionality. However, this expansion also increases vulnerabilities, exposing IoT networks to increasingly sophisticated cyberattacks.  Intrusion Detection Systems (IDS) are crucial for mitigating these threats, and recent advancements in Machine Learning (ML) offer promising avenues for improvement. This research explores a hybrid approach, combining several standalone ML models such as Random Forest (RF), XGBoost, K-Nearest Neighbors (KNN), and AdaBoost, in a voting-based hybrid classifier for effective IoT intrusion detection. This ensemble method leverages the strengths of individual algorithms to enhance accuracy and address challenges related to data complexity and scalability.  Using the widely-cited IoT-23 dataset, a prominent benchmark in IoT cybersecurity research, we evaluate our hybrid classifiers for both binary and multi-class intrusion detection problems, ensuring a fair comparison with existing literature.  Results demonstrate that our proposed hybrid models, designed for robustness and scalability, outperform standalone approaches in IoT environments. This work contributes to the development of advanced, intelligent IDS frameworks capable of addressing evolving cyber threats.
\end{abstract}


\section{Introduction}
\IEEEPARstart{T}{he} Internet of Things (IoT) has brought about new dimensions in the modern world, facilitating connection between devices through networks in fields such as healthcare, transportation, and smart cities. Among all these advantages IoT technology brings, cyber-crime attackers readily attack critical vulnerabilities. This is because IoT networks are extensive in scale and, therefore, complex, making them hot targets for attacks related to data breaches, malware infections, and Distributed Denial-of-Service (DDoS) incidents. Although these threats are becoming more sophisticated, there is a need for effective and adaptive intrusion detection mechanisms to cope with the ever-expanding complexity of cyberattacks.

This development has opened new avenues to improve IoT security through computational advancements. Learning-based methods are generally effective in detecting malicious activities. In contrast to static and rule-based systems, they can analyze large volumes of data to reveal hidden patterns and detect abnormalities in real-time. Their advantage is their adaptability, as the system can always stay one step ahead of the bad guys. These approaches equip IoT networks with dynamic defense against cyber risks by drawing on insights gained from various data sources.

\IEEEpubidadjcol

Intrusion Detection Systems (IDS) play a critical role in cybersecurity by identifying malicious activities in a network or system. In the recent past, researchers often employed machine learning (ML) and deep learning (DL) methods, either as standalone models or in hybrid configurations, to enhance IDS capabilities, especially to detect anomalies (abnormal behaviors or activities within their systems or networks). These methods cater to binary classification (e.g., distinguishing between malicious and benign activities) and multi-class classification (e.g., identifying specific types of attacks).

In this study, we utilize IoT-23 Dataset, a benchmark IoT dataset for intrusion detection research, developed by the Stratosphere Laboratory~\cite{garcia2020iot}. The main focus of this dataset is on malicious and benign traffic. The dataset contains 23 labeled scenarios and a wide range of attacks like DDoS, data exfiltration, malware activities, and normal traffic from IoT devices such as smart cameras, motion sensors, and smart hubs. It is captured in packet capture format, hence suitable for signature-based and ML based analysis. This dataset is of great value in evaluating IDS due to its representation of real-world IoT scenarios, the balance between attack and benign data, and the coverage of different attack vectors in IoT networks.

Early steps of our study were presented in \cite{akif2024harnessing}, especially involving binary classification with standalone ML methods. In this study, we explore a hybrid approach that combines the strengths of multiple computational models to improve intrusion detection in IoT systems. This approach addresses common challenges, such as balancing detection accuracy with scalability and minimizing false alarms. By uniting complementary techniques, the framework aims to create a more resilient solution capable of countering previously unseen threats. Through this work, we hope to offer practical strategies for building secure IoT environments tailored to the unique demands of this rapidly growing field.


The paper's organization is as follows: Section~\ref{sec:Section2} provides the related work (literature) in the field. Section~\ref{sec:Section3} includes the prominent phase of this study, which is the ``Data Pre-Processing''. Section~\ref{sec:Section4} provides the foundations and methodology related to ML algorithms used in this study. Section~\ref{sec:Section5} plots all the necessary experimental results for evaluating the proposed Mix-Zone scheme for the IoBT. Finally, the overall work is deduced and concluded in Section~\ref{sec:Section6}.

\section{Related Work}\label{sec:Section2}
Given their growing and linked character, IoT networks—which are increasingly vulnerable to cyber-attacks—rely on IDS to maintain their security. Using both binary and multi-class intrusion detection models, several machine ML and DL methods have been employed to improve IDS. This review of the literature investigates several current works addressing intrusion detection problems using various techniques, both as stand-alone models and in hybrid setups.

The authors of \cite{bhandari_distributed_2023} presented a new AI framework that detects malware in IoT devices to mitigate cyber-attacks. The authors focus on enhancing security in various use cases for smart environments through an all-inclusive AI-enabled approach in this paper. Emulation of a smart environment employing the Raspberry Pi and NVIDIA Jetson as gateways in capturing data from IoT devices connected via the MQTT protocol, therefore enabling monitoring of real-time malware attacks for their prediction. In this work, many models of AI have been evaluated, among which the DNN model demonstrated superior accuracy and classification capability with an F1-score of 92\% and detection accuracy of 93\% on Edge-IIoTset and IoT-23. Concerns about the impact on system resources by specifying metrics are drawn to traffic and CPU usage on both devices, while challenges include the lack of ground-truth data in most cyberattacks. Future research shall 
be on few-shot learning, lightweight model implementation, DL cutting-edge methodologies, penetration testing, and the use of additional sensor and actuator data to enhance the anomaly detection system.

Using the IoT 2023 dataset as a thorough benchmark, the study of \cite{sarobin2024comparative} tackles the issue of feature extraction from IoT data. The goal is to gain a better understanding of the dataset's properties and possible uses by evaluating both classic statistical approaches and ML-based methods. Feature extraction takes on more significance in the context of the IoT because of the "curse of dimensionality," the well-known fact that data processing and analysis get more complicated as the number of dimensions grows. Various techniques have been surveyed to place them in the context of their strengths in capturing relevant information, reducing dimensionality, and improving performance in IoT analytics. Some key findings in this respect include the Hughes phenomenon: classifier performance may get better with more features up to some optimal point before deteriorating. This paper guides the choice of suitable feature extraction methods to be deployed for various IoT applications via ample experiments and performance analysis. This will, therefore, help in the practical development of IoT solutions in 2023 and beyond. Besides, according to the authors, little effect of reducing features on the model performance is up to an accuracy of 93.04\% using Decision Trees and 93.05\% using Random Forest (RF) models. 


Using ML approaches, Prazeres \textit{et al.} (2022) evaluates AI-based malware detection in IoT network traffic. The study makes use of the IoT-23 dataset using real IoT network traffic of both benign and malicious, including numerous forms of malware including botnets and DDoS attacks. Key strategies to categorize network traffic are feature selection, data normalization, and the application of several ML models (Logistic Regression, RF, ANN, and Naïve Bayes) \cite{prazeres_evaluation_2022}. 


Especially in multi-class issues, RF has repeatedly shown to be a useful classifier in identifying intrusions. For example, with high accuracy rates in multi-class classification, a RF application to the IoT-23 dataset revealed better performance than other ML models~\cite{malele2023testing}.

Likewise, XGBoost, known for its boosting power, has demonstrated extraordinary intrusion detection capability. As network IDS built for IoT networks show, XGBoost lowers mistakes and raises classification accuracy by iteratively strengthening weak models \cite{banadaki2020evaluating}. Research also shows how well it can handle prevalent network dataset class imbalance problems \cite{jiang2020network}.



Using multiclassification models within the PySpark architecture, a 2024 study by Alrefaei et al. \cite{alrefaei_using_2024} offers an IoT network real-time intrusion detection system (IDS). One-Vs- Rest (OVR) method ML approaches include RF, Decision Trees, Logistic Regression, and XGBoost help to enhance detection accuracy and minimize prediction latency. Class imbalance is solved via data cleansing, scaling, and SMote using the IoT-23 dataset. RF displayed the fastest prediction time at 0.0311 seconds, but XGBoost achieved the highest accuracy at 98.89\%, underlining the system's value in real-time IoT threat detection and so reducing security concerns.

Combining the advantages of several classifiers, ensemble approaches have shown notable gains in IDS performance. In particular, hybrid models, which combine classifiers in voting systems to improve overall accuracy and detection rate, have encouraging results. Upadhyay et al. (2021) presented a majority voting ensemble combining RF, XGBoost, and KNN with additional classifiers for SCADA-based power grids. Their model demonstrated gains in binary and multi-class classification by selecting features using Recursive Feature Elimination and then utilizing majority voting to increase precision and recall \cite{upadhyay2021intrusion}.


A hybrid model that included RF, XGBoost, and KNN and was improved using feature selection approaches was also investigated by Liu et al. (2023). When tested on many datasets, their approach showed an increase in overall detection accuracy and a decrease in false positives. In this study, the voting system of a hybrid classifier, which combines other individual classifiers, is achieving better performance than stand-alone classifiers \cite{liu2023hybrid}.


Research shows that a voting ensemble of classifiers including RF, XGBoost, AdaBoost, KNN, and SVC performs better than any classifier taken on alone. Leevy et al. (2021) evaluated different ensemble models, including XGBoost and RF, to find assaults in the framework of the IoT. Their studies show that in terms of adaptation and accuracy, ensemble models usually outperform individual classifiers \cite{leevy2021detecting}.


When testing IDS, the gold standard is the IoT-23 dataset, which simulates real-world IoT network traffic. It provides several attack scenarios that researchers can use to test the multi-class classification capabilities of DL and ML models. Detection rates surpass 99\% when models based on RF and CNN are used \cite{derhab2020intrusion}.

ElKashlan \textit{et al.} proposed a ML-based IDS to safeguard electric vehicle charging stations (EVCSs) in an IoT environment. They evaluated many machine-learning techniques to identify fraudulent traffic in both binary and multiclass traffic models using a real-world IoT dataset.  The suggested solution seeks to improve EVCS security by successfully detecting and thwarting cyberattacks, guaranteeing the ecosystem's stability, and guarding against possible interruptions to day-to-day operations.  Along with defining the EV charging ecosystem, the article looks at the many attack types that could happen and the vulnerabilities that each component faces. The traffic and possible assaults on an EVCS were represented by them using a native IoT dataset, IoT-23, which is derived from actual IoT devices \cite{elkashlan_machine_2023}.

To improve intrusion detection in IoT networks, Ullah et al. (2021) suggested a hybrid model that combines Convolutional Neural Networks (CNN) with Gated Recurrent Units (GRU). To accomplish reliable binary and multiclass classifications, it applies sophisticated preprocessing and feature selection algorithms to a variety of datasets (BoT-IoT, MQTT-IoT-IDS2020, etc.). The model outperforms earlier techniques with high accuracy (up to 99.96\%) and efficient detection rates across a range of cyberattack types. Its architecture uses strategies like dropout layers and early halting to overcome common problems like overfitting and data imbalance. In order to improve detection skills, future research areas will use Generative Adversarial Networks (GANs) \cite{ullah_towards_2021}. 

Using generative adversarial networks (GANs), Park et al. suggest an improved AI-based network intrusion detection system (NIDS). To identify anomalies and learn the typical behavior of network traffic, the system makes use of a DL model.  In order to increase the NIDS's accuracy, the study also creates fake data using GANs.  When tested on a benchmark dataset, the suggested system outperforms the most advanced NIDSs, according to the results.  The study comes to the conclusion that the suggested approach can be utilized to increase computer network security and is efficient at identifying network intrusions \cite{park_enhanced_2023}.

To identify cyberattacks in IoT networks, Abdalgawad et al. investigated the use of generative DL models, such as Adversarial Autoencoders (AAE) and Bidirectional Generative Adversarial Networks (BiGAN). They analyzed the IoT-23 dataset, which includes network traffic information from actual IoT devices, such as Amazon Echoes and smart door locks. These models were trained to recognize a variety of attacks, such as botnet activity from Mirai, Okiru, and Torii, as well as DDoS attacks. According to the study, these generative models outperformed more conventional ML techniques like RF in terms of intrusion detection. This study shows how generative DL can be used to enhance cybersecurity in the expanding IoT environment \cite{abdalgawad_generative_2022}.


\section{Data Pre-Processing}\label{sec:Section3}
Data pre-processing is essential in preparing datasets for analysis in training ML models. This stage ensures that the data is well organized, cleaned, and structured in such a way as to guarantee meaningful insights in the subsequent processing. Data pre-processing includes identifying the dataset type, understanding class distributions, uncovering patterns, and extracting other useful information that will drive how the data should be analyzed. This leads to improved model performance and reliability~\cite{ahmad2019data}.

The following sections will discuss the steps to pre-process our data.

\subsection{Data Analysis}
Data analysis is a crucial stage in the pipeline of data pre-processing. It is used to study the dataset for its structure, distribution of classes, and any inherent patterns or anomalies. These are important for formulating a strategy on how the data should be transformed, normalized, or cleaned to optimize the results in further analysis.

\subsubsection*{Class Distribution}
We used 100,000 data observations for binary classification, and for multi-class classification, we used 69,398 data observations. Each class has 50,000 observations for binary classification and for multi-class classification, each class has around 10,000 observations. 


For binary classification there are only two classes: Benign and Malicious and for the multi-class classification, there are seven classes: Benign, C\&C- HeartBeat, DDoS, Okiru, PartOfHorizontalPortscan, C\&C, and Attack. 



\subsection{Missing data Handling}
Missing data handling refers to missing values so they don't negatively impact target prediction. To address the missing values in the IoT-23 dataset, we substituted the empty values of numerical features with the ``mean'' value of that feature. We have chosen to fill the empty values with mean because if we keep them, it will compromise the data integrity and distort the proper data distribution. Additionally, there was a categorical feature called service that had been filled with the value ``unknown'' and also had empty values. 

We didn’t use the ``mean'' values for the categorical features because this feature will be converted into 0 and 1 later. If we fill the empty values of this feature with the mean, then it cannot indicate the classes of that feature because the mean will be the average of that feature, which is a float value. Effective handling of missing data achieves the integrity of model training and minimizes the risk of biased or incorrect predictions \cite{raj2023machine}.



\subsection{Categorical Feature Conversion}
We need to convert the categorical features into numeric values to train the ML models because the models cannot understand a word as an input value. To do that, we applied the One-hot encoding method, which converts features such as if a specific class is present in an observation. It denotes the value of that observation for that class as one otherwise 0.

\subsection{Feature Engineering}
The goal of feature engineering in ML is to improve a model's prediction power by enhancing it with additional features or tweaking its current features. It converts raw data into a format better suited for model training so the computer can more effectively find significant trends, correlations, and patterns.

Practical feature engineering uses domain expertise; for instance, dimensionality needs to be decreased to increase computational efficiency, and data quality needs to be enhanced by building meaningful features. This technique makes model accuracy and interpretability and tailoring ML solutions to particular application contexts possible through this technique~\cite{katya2023exploring}.

\begin{figure}[h!]
    \centering
    \includegraphics[width=0.99\columnwidth]{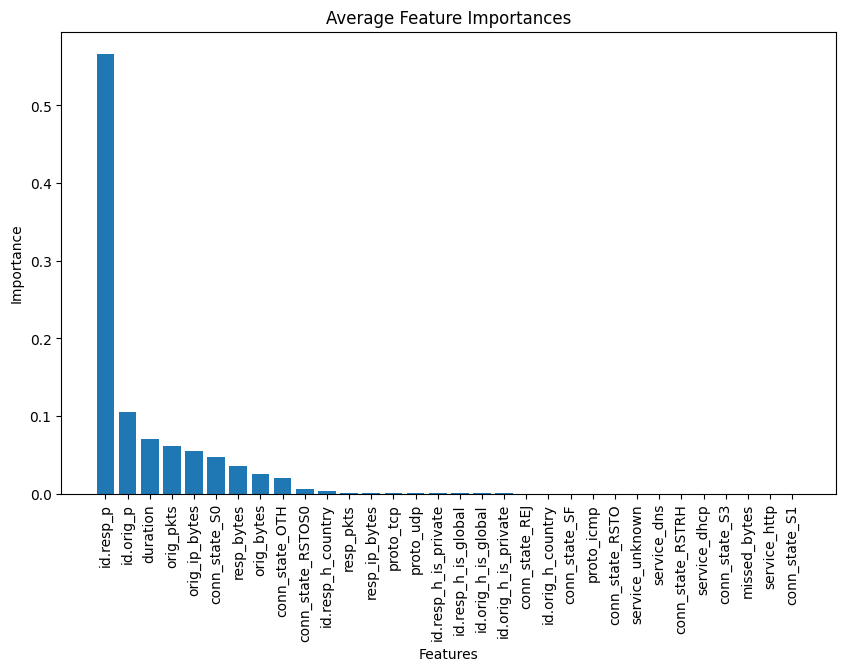}
    \caption{Feature importance of binary classification}\label{Fig.X}
\end{figure}

For the dataset, first, we extracted information from two of the features containing the IP addresses. The first information we extracted tells us if the IP is private or global, and the second information tells us the countries where the IP addresses are located. After that, we removed those IP address columns because they would overfit the model with training data, which is unsuitable for our model. Some more ID columns are also being removed.

After removing the ID columns, a categorical feature named ``history'' increased the dimensionality problem. For this, the model will become more complex and may capture noise rather than the actual underlying pattern of the dataset, which will cause the model to overfit. So, to mitigate this problem, we also removed that feature. After the above tasks, we did the feature importance test which will decrease the complexity of the dataset and also it will make the dataset more generalized because by doing this test we will select the topmost impactful features for our models and will remove the remaining features.

Figure~\ref{Fig.X} shows the importance of every feature for binary classification and from that graph, we can say that the top 11 features have the most impact on predicting the target because those features have an importance score greater than 0 but the rest of it is very close to 0 so, we had kept the first 11 features and removed the rest from our dataset.

\begin{figure}[h!]
    \centering
    \includegraphics[width=0.99\columnwidth]{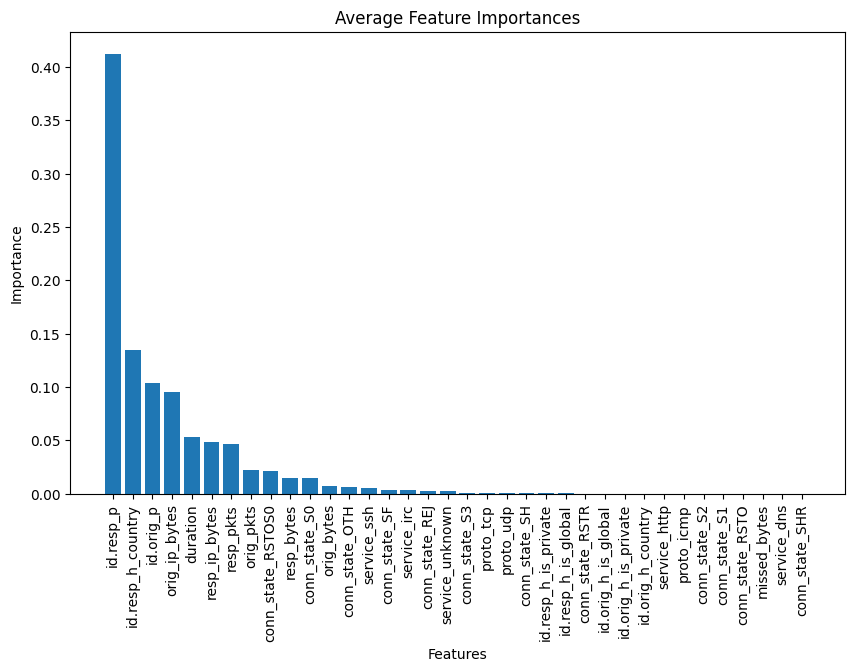}
    \caption{Feature importance of multi-class classification}\label{Fig.Xx}
\end{figure}

Figure~\ref{Fig.Xx} shows the importance of every feature for multi-class classification and from that graph, we can say that the top 18 features are impacting the most to predict the target because those features have an importance score greater than 0 but the rest of it is very close to 0 so, we had kept the first 18 features and removed the rest from our dataset.


\subsection{Data Splitting}
In ML, data splitting refers to partitioning a dataset into distinct subsets to facilitate model training, validation, and evaluation. This step is critical for ensuring that an ML model does not simply memorize the training data, a problem known as overfitting, but instead learns to generalize effectively to unseen, real-world data. Proper data splitting helps evaluate the model's performance on independent data, objectively assessing its predictive capabilities. The following sections discuss the specific data-splitting ratios used in our models and their significance for achieving robust generalization.

\subsubsection{Splitting Ratio of the models}
We had taken 80\% of the total data for training and 20\% of the total data for testing. As we use cross-validation, which we will present later, it will randomly split the training dataset into validation and training data. 


\subsubsection{K-Fold Cross Validation}
K-fold cross-validation is a resampling method for assessing ML models that involves splitting the data into k equal-sized folds. Each iteration's validation set is one-fold, while the training set consists of the remaining K-1 folds. A more accurate evaluation of the model's efficacy can be obtained by averaging the results of this k-times method.

It aids in avoiding overfitting. Moreover, it offers a more reasonable assessment of a model's generalization ability.
Picking K (usually 5 or 10) strikes a compromise between computing expense and the stability of the performance estimate. With K-fold cross-validation, you can get strong accuracy metrics from more enormous datasets without collecting extra validation data~\cite{xiong2020evaluating}.

\subsection{Feature Scaling}
In ML, feature scaling is a preprocessing method used to change the range or distribution of data values, guaranteeing consistency across features. For KNN, this stage is essential since it standardizes or normalizes the data to a uniform scale, lowering the variability among features.
Eliminating bias resulting from features of different magnitudes, which could skew model predictions, depends on feature scaling. By lowering variations in feature range, techniques such as normalization and standardization help ML models to operate better~\cite{ozsahin2022impact}. For the feature scaling of the IoT-23 dataset, we have used Min-Max Scaler, as provided below.

\subsubsection*{Min-Max Scaling}
Min-Max in ML scaling is a normalizing technique used to rescale traits within a specific range, usually [0, 1]. This approach guarantees that no feature dominates the others and contributes equally~\cite{ambarwari2020analysis}. It is also beneficial in the cases of variable scales or units for the features. The conversion is accomplished with the following formula:
\begin{equation}
X' = \frac{X - X_{\text{min}}}{X_{\text{max}} - X_{\text{min}}}
\end{equation}
where \( X' \) is the value after scaling, \( X \) is the original value of that particular observation, \( X_{\text{min}} \) is the minimum feature value, and \( X_{\text{max}} \) is the maximum feature value.

\subsection{Challenges}
During the pre-processing of the data, we faced three crucial challenges, which were data volume, class imbalance, and information leakage:

\subsubsection{Data Volume}
Data volume plays a vital role in ML. Our IoT-23 dataset had 21 GB of data with around 60 million observations, making the pre-processing tasks like feature scaling, one hot encoding, and label encoding very slow. Moreover, it made the training process very slow, too, and that’s why, to mitigate these problems, we took a portion of data for binary and multi-class classification, which will not affect the model performance negatively. Also, after doing that, the pre-processing tasks became very fast along with the training process.

\subsubsection{Class Imbalance}
In ML, the class imbalance results from an unbalanced distribution of classes in a dataset whereby one class (or a small number of classes) has significantly more instances than others. This usually happens in classification problems when some events or results are rare in relation to others. There was a class imbalance in our dataset consisting of eleven assault kinds or classes; as a result, we removed the classes with the fewest observations and kept seven that had a sufficient amount of data.

\subsubsection{Information Leakage}
In ML, information leakage results from a model inadvertently accessing unrestricted material from the training set, producing poor generalization and unrealistic performance measures and possibly undermining the predicting capability of the model. In our case, after the data pre-processing, when the training of the models just started, we have observed that from the very first epoch, the accuracy is too high, like over 90\%, which is not sophisticated because in the very first epoch, no model can be perfect or close to perfect so after getting this kind of output, we had suspected that there is an information leakage which was during the data scaling. Before we knew the problem, we scaled the whole dataset. At that time, the scaler we had used gained access to all the data, which was not supposed to have happened, and we scaled the data according to that. After finding the problem, we split the data. Then, we fit our scaler on the training data and used that scaler to transform our training, testing, and validation dataset. In this way, the scaler was not gaining access to the validation and testing data.

\section{Methodologies}\label{sec:Section4}
\subsection{Binary vs. Multi-class Classification}

Binary classification represents predicting if an IoT device is being attacked or not which means it has two outputs one is benign and another one is malicious if the device is being attacked it will output as malicious and if it is not, it will output as benign. We had done research work on Binary Classification \cite{akif2024harnessing} where we had used several ML and DL techniques and among those our XGBoost has performed the best with 98.9\% Accuracy, 98.5\% of Precision, 98.7\% of Recall and 98.9\% of F1-Score.

Multi-class classification represents predicting the attack type like Okiru, DDoS, etc., and if it is not being attacked then representing it as benign.

We proposed two different hybrid models that use voting for classification. In these two models, we combined XGBoost, RF, and KNN for binary classification and combined RF, XGBoost, and AdaBoost for multi-class classification. In the next sections, we will present these methods.

\subsection{Random Forest}
Random Forest (RF) is one of several ensemble methods of ML that offer a versatile approach suitable for regression and classification. The mode of the classes used for classification tasks can be generated by training with a large number of decision trees. This method makes RF quite successful for classification and regression projects and helps lower overfitting~\cite{schonlau_random_2020}.

\subsection{Extreme Gradient Boosting (XGBoost)}
Considered a subset of ensemble learning and a ML method included in the gradient boosting framework is extreme gradient boost boosting (XGBoost). Beginning with simple learners, decision tree regularization methods help increase model generalization. XGBoost is computationally efficient; it offers sound processing, perceptive feature significance analysis, and smooth management of missing data. In a study by Chen et al. ~\cite{chen_improved_2020}, the following statement is mentioned: "Especially in problems with complicated data linkages, XGBoost can reach state-of-the-art performance using this sequential boosting with regularization approach."

\subsection{K-Nearest Neighbors (KNN)}
Using a distance metric, typically Euclidean distance, the KNN algorithm classifies data points according to the majority label of their nearest neighbors; it finds extensive use in non-parametric ML tasks like regression and classification. Although KNN is computationally heavy for big datasets since it must compute distances for every query point, it is popular for its ease of use and effectiveness in uses, including pattern detection and recommendation systems~\cite{wang_improved_2021}.

\subsection{AdaBoost}
Combining several weak classifiers, the AdaBoost (adaptive boost) algorithm is an ensemble learning method that produces a strong classifier with better accuracy. AdaBoost strengthens the general prediction performance of the model by iteratively changing the weights of training data, hence focusing on the samples that past classifiers misclassified. This method especially helps to raise classification accuracy in many different fields~\cite{li_improved_2023}.

\subsection{Hybrid Voting Classifier}
In ML, a hybrid model combines several methods or models to improve predicted accuracy and robustness based on voting. In these hybrid systems, voting functions as a consensus process whereby every model votes on the forecast result, and the final choice is determined depending on a weighted or majority voting system. Particularly useful in classification problems, voting-based hybrid algorithms use the different capabilities of several models to manage heterogeneous and complex data, hence improving prediction accuracy~\cite{du_research_2022}.

\begin{figure}[h!]
    \centering
    \includegraphics[width=0.99\columnwidth]{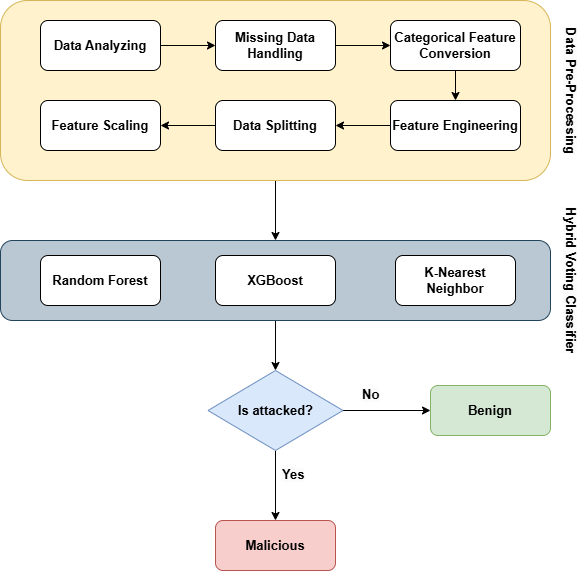}
    \caption{Flow Chart Representation of Our Proposed Hybrid Model Algorithm for Binary Classification}\label{Fig.4a}
\end{figure}

\begin{figure}[h!]
    \centering
    \includegraphics[width=0.99\columnwidth]{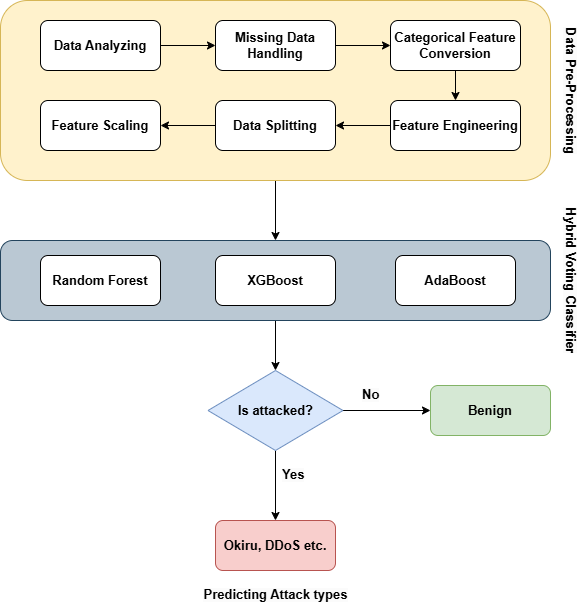}
    \caption{Flow Chart Representation of Our Proposed Hybrid Model Algorithm for Multi-Class Classification}\label{Fig.4b}
\end{figure}

Below (Figures \ref{Fig.4a} and \ref{Fig.4b})  are the flow chart representation of the decision process in our Binary and Multi-class classification models:

In Figure \ref{Fig.4a} one can observe the flow chart of Binary classification and can find out that the pre-processing part is a combination of different steps like Data Analyzing, Missing Data Handling, Categorical Feature Conversion, Feature Engineering, Data Splitting and Feature Scaling. One can also see that for binary classification RF, XGBoost, and KNN. It can also be seen that this model predicts if the device is Benign or Malicious.

In Figure \ref{Fig.4b} it can be found that the pre-processing steps of multi-class classification are the same as binary classification. Still, the models that had been combined inside of the hybrid voting classifier are different which are RF, XGBoost, and AdaBoost. Also during the decision making the model predicts whether it is benign or what type of attack it is.

\section{Evaluation, Results and Discussion}\label{sec:Section5}
We have evaluated our models in terms of 4 different scales; F1-Score, Accuracy, Precision, and Recall. Below are the formulas for those metrics:


\begin{align*}
\text{Accuracy} & = \frac{\text{Number of Correct Predictions (TP + TN)}}{\text{Total Number of Predictions (TP+TN+FP+FN)}} \\
\end{align*}




\[
\text{Precision} = \frac{\text{TP}}{\text{TP} + \text{FP}},  
\text{Recall} = \frac{\text{TP}}{\text{TP} + \text{FN}}
\]




\[
\text{F1-Score} = 2 \times \frac{\text{Precision} \times \text{Recall}}{\text{Precision} + \text{Recall}}
\]

Here; TP stands for True Positive, TN stands for True Negative, FP stands for False Positive and FN stands for False Negative.



We had listed down the evaluation metrics values from the models of our related works in a tabular format where Table~\ref{Table.1} is for Binary Classification and Table~\ref{Table.2} is for Multi-Class Classification.

After analyzing the comparison tables (Tables ~\ref{Table.1}-~\ref{Table.2}), we can conclude that our hybrid model for binary classification outperforms all other relevant proposed schemes and models in the literature, in all metrics, Accuracy (A), Precision (P), Recall (R), F1-Score (F1), respectively. For multi-class classification, our hybrid model outperforms all other relevant proposed schemes and models in the literature in all metrics (A, P, R, F1); except RF of Paper \cite{prazeres_evaluation_2022} (slightly worse in P and F1), Filtered Classifier of \cite{elkashlan_machine_2023} (slightly worse in A). However, these models are not tree-based, therefore they are not expected to be as scalable and faster as our proposed Hybrid model, which is based on tree-based ML algorithms.

Now, the confusion matrix of the hybrid models of binary and multi-class classification will be described below:


\begin{figure}[h!]
    \centering
     \includegraphics[width=0.79\columnwidth]{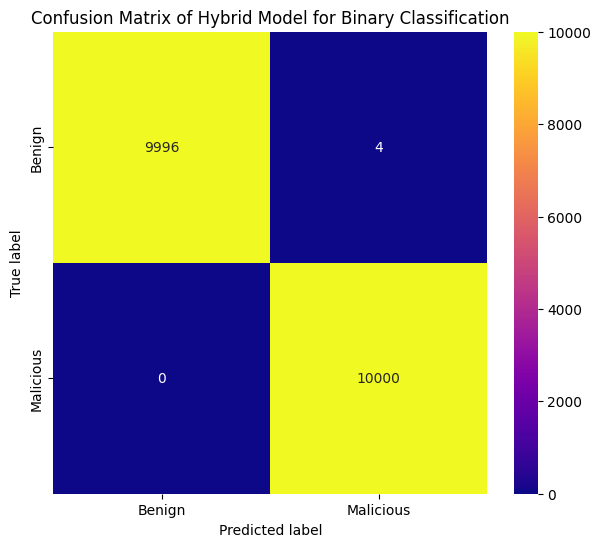}
    \caption{Confusion Matrix of Binary Classification}\label{Fig.5}
\end{figure}


\begin{figure}[h!]
    \centering
    \includegraphics[width=0.99\columnwidth]{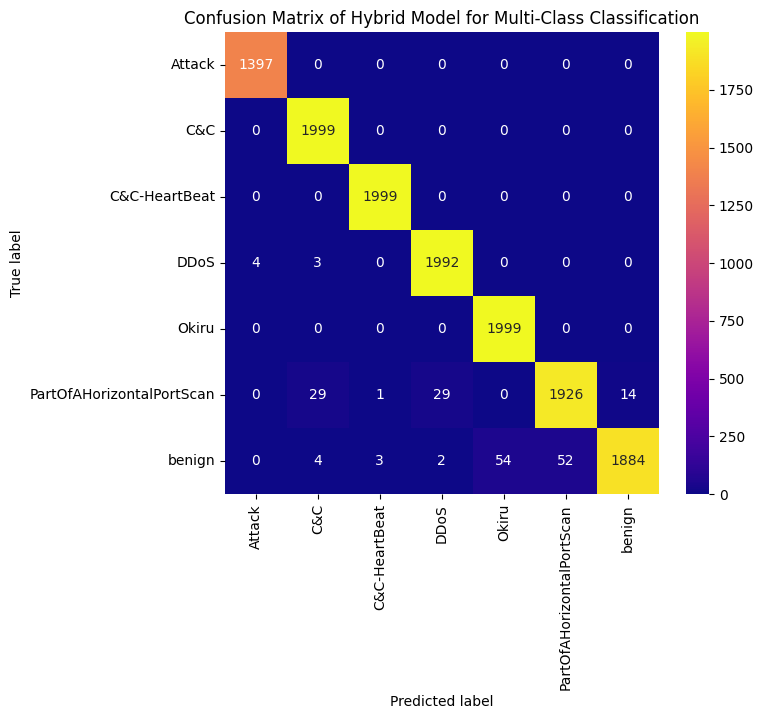}
    \caption{Confusion Matrix of Multi-Class Classification}\label{Fig.6}
\end{figure}

From the confusion matrix for binary classification of our hybrid model using the testing dataset provided in Fig.~\ref{Fig.5}, some valuable insights can be extracted. First of all, majority of the predictions are correct (9,996 for benign and 10,000 for malicious) for both of the classes which means the model is not biased for a specific class. Moreover, we have no FN (type 2 error) which is 0 and as the model's main target is to detect attacks correctly and it is doing it perfectly, this model will be very much suitable in security applications which is a crucial success for us. Also, no FN led us to very R. Furthermore, the model's FP (type 1 error) is 4 which led us to a very high P.

On the other hand, Fig.~\ref{Fig.6} which is the confusion matrix for multi-class classification of our hybrid model using the testing dataset, also gives us some valuable insights. First of all, most of the predictions are aligned correctly which is indicated by the diagonal values. It also tells us that the model is not biased. Moreover, if we analyze class-wise performance then we can see Attack, C\&C, C\&C-HeartBeat, DDoS, and Okiru are almost perfectly predicted and PartOfAHorizontalPortScan has the highest misclassification among the other attack types which is 73 but this number is very low because only 3.6\% of the total number of observations for that class are being predicted incorrectly. Furthermore, the average FN (type 2 error) and FP (type 1 error) are very low, which is 0.74\% and 0.37\%, respectively, which indicates that as a multi-class classification model, it is able to predict multiple attack types almost perfectly and is very much suitable for IDS in IoT.


Now, we will elaborate on the accuracy curves of these two hybrid models:

\begin{figure}[h!]
    \centering
    \includegraphics[width=0.97\columnwidth]{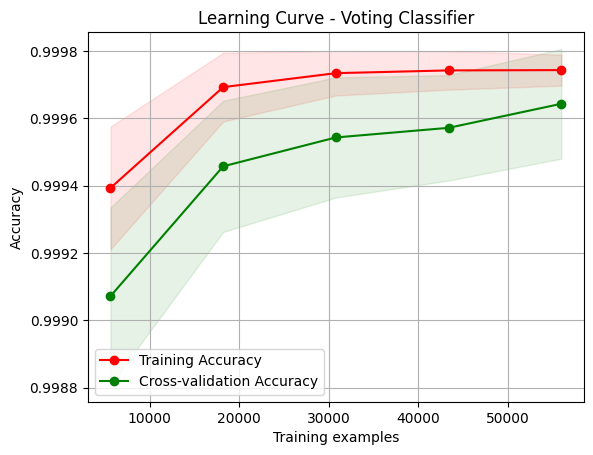} %
    \caption{Accuracy Curve for Binary Classification}\label{Fig.7}
\end{figure}

The accuracy curve provided in Fig.~\ref{Fig.7} for binary classification shows the training and cross-validation accuracy curve using the training dataset. From the curve, it can be said that the training and cross-validation accuracy started from around 99.94\% and 99.9\% respectively and it gradually increased to around 99.98\% for training and 99.97\% for cross-validation accuracy. The graph also indicates no spike, so the validation and training curves converge smoothly. Moreover, the validation accuracy curve is not going downward at any point, so there is no sign of model overfitting, which is excellent!

\begin{figure}[h!]
    \centering
    \includegraphics[width=0.9\columnwidth]{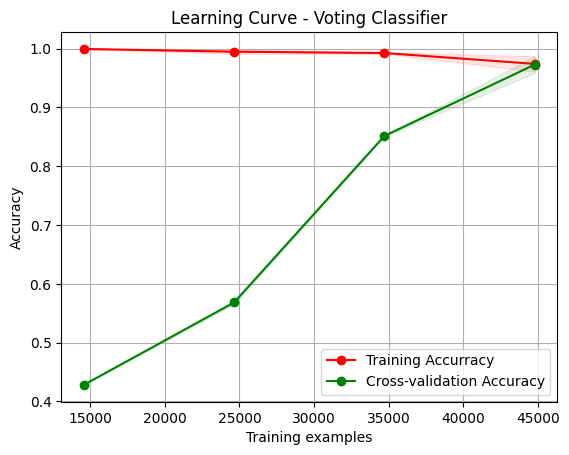}
    \caption{Accuracy Curve for Multi-Class Classification}\label{Fig.8}
\end{figure}

The accuracy curve provided in Fig.~\ref{Fig.8} for the multi-class classification shows the training and cross-validation accuracy curve using the training dataset. From the curve, it can be said that the cross-validation accuracy started from around 41\%, but the training accuracy started from 99.9\%. Gradually, after so many training iterations, the cross-validation accuracy slowly climbed to the accuracy of around 99\% but training accuracy was between 99\% and 99.9\%. The graph also shows no spike in the validation or training curve, so it is converging smoothly, also, there is no point when the validation accuracy started to decrease compared to the training accuracy so there is no sign of model overfitting, which is also great!

\begin{table*}[h]
\centering
\caption{Comparison Table of the Evaluation Metrics for Binary Classification}\label{Table.1}
\begin{tabular}{|l|l|c|c|c|c|}
\hline
\textbf{Model Name} & \textbf{Reference Paper} & \textbf{Testing Accuracy} & \textbf{Precision} & \textbf{Recall} & \textbf{F1 Score} \\ \hline

XG Boost & Our earlier paper \cite{akif2024harnessing} & 98.9 & 98.5 & 98.7 & 98.9 \\ \hline

LR &  Paper \cite{prazeres_evaluation_2022}& 91.99$^a$ & 85.11 & 88.90 & 91.94$^a$ \\ 
RF &  Paper \cite{prazeres_evaluation_2022}& 99.98$^a$ & 99.95 & 99.95 & 99.95$^a$ \\ 
NB &  Paper \cite{prazeres_evaluation_2022}& 96.14$^a$ & 92.45 & 98.05 & 95.17$^a$ \\ 
ANN &  Paper \cite{prazeres_evaluation_2022}& 99.79$^a$ & 99.57 & 99.70 & 99.64$^a$ \\ \hline

Naïve Bayes & Paper \cite{elkashlan_machine_2023} & 86.7 &  85.50$^b$ & 77.50$^b$ & 80.00$^b$ \\ 
J48 & Paper \cite{elkashlan_machine_2023} & 97.4 & 95.50$^b$ & 97.50$^b$ & 96.50$^b$ \\ 
Attribute select & Paper \cite{elkashlan_machine_2023} & 91.5 & 87.00$^b$ &  93.00$^b$&  89.50$^b$\\ 
Filtered Classifier & Paper \cite{elkashlan_machine_2023} & 99.2 & 98.50$^b$& 99.00$^b$&  98.50$^b$\\ \hline

Hybrid (CNNGRU) & Paper \cite{ullah_towards_2021} & 99.96 & 99.90 & 99.95 & 99.93 \\ \hline

\textbf{Hybrid (XGBoost, RF and KNN)} & \textbf{Ours} & \textbf{99.99} & \textbf{99.99} & \textbf{99.99} & \textbf{99.99} \\ \hline
\end{tabular}%
\label{tab:binary_classification}
\\
\footnotesize{
$^a$: Calculated by using the precision and recall of the values at the binary classes provided in~\cite{prazeres_evaluation_2022}.\\
$^b$: Calculated by using the average of the values at normal and attack cases provided in~\cite{elkashlan_machine_2023}.\\

}
\end{table*}

\begin{table*}[h]
\centering
\caption{Comparison Table of Evaluation Metrics for Multi-Class Classification}
\label{Table.2}
\begin{tabular}{|l|l|c|c|c|c|}
\hline
\textbf{Model Name} & \textbf{Reference Paper} & \textbf{Testing Accuracy} & \textbf{Precision} & \textbf{Recall} & \textbf{F1 Score} \\ 
\hline
DNN & Paper \cite{bhandari_distributed_2023} & 93 & 97 & 92 & 94 \\ 
SVM and GB & Paper \cite{bhandari_distributed_2023} & 95 & 54 & 43 & 48 \\ 
RF & Paper \cite{bhandari_distributed_2023} & 95 & 59 & 44 & 50 \\ 
DT & Paper \cite{bhandari_distributed_2023} & 95 & 56 & 45 & 50 \\ 
NB & Paper \cite{bhandari_distributed_2023} & 82 & 29 & 49 & 36 \\ \hline

LR &  Paper \cite{prazeres_evaluation_2022}& 84.73$^a$ & 91.54$^a$ & 91.93$^a$ & 91.50$^a$ \\ 
RF &  Paper \cite{prazeres_evaluation_2022}& 98.55$^a$ & \textbf{99.83}$^a$ & 98.72$^a$ & \textbf{99.25}$^a$ \\ 
NB &  Paper \cite{prazeres_evaluation_2022}& 56.59$^a$ & 63.12$^a$ & 84.55$^a$ & 67.74$^a$ \\ 
ANN &  Paper \cite{prazeres_evaluation_2022}& 74.76$^a$ & 90.24$^a$ & 81.34$^a$ & 84.70$^a$ \\ \hline

XG Boost & Paper \cite{alrefaei_using_2024} & 98.89 & 98.93 & 98.89 & 98.89 \\ \hline

Naïve Bayes & Paper \cite{elkashlan_machine_2023} & 77 & 92$^b$ & 68$^b$ & 68$^b$ \\ 
J48 & Paper \cite{elkashlan_machine_2023} & \textbf{99.2} & 98$^b$ & 99$^b$ & 98$^b$ \\ 
Attribute select & Paper \cite{elkashlan_machine_2023} & 97.12 & 94$^b$ & 98$^b$ & 96$^b$ \\ 
Filtered Classifier & Paper \cite{elkashlan_machine_2023} & \textbf{99.2} & 98$^b$ & \textbf{99}$^b$ & \textbf{99}$^b$ \\ \hline

Hybrid (CNNGRU) &  Paper \cite{ullah_towards_2021} & 95.81$^c$ & 96.79$^c$ & 98.96$^c$ & 97.77$^c$ \\ \hline

DNN & Paper \cite{park_enhanced_2023}  & 93.10 & 94.57$^d$ & 82.33$^d$ & 85.00$^d$ \\ 
CNN, DNN$_{AE}$, and CNN$_{AE}$ & Paper \cite{park_enhanced_2023} & 93.70 & 95.47$^d$ & 82.80$^d$ & 85.97$^d$ \\ 
LSTM & Paper \cite{park_enhanced_2023} & 93.50 & 95.13$^d$ & 82.80$^d$ & 85.77$^d$ \\ 
Hybrid (G-LSTM, DNN$_{AE}$, CNN$_{AE}$) & Paper \cite{park_enhanced_2023} & 95.9 & 96.80$^d$ & 93.33$^d$ & 94.63$^d$ \\ \hline

KNN &  Paper \cite{abdalgawad_generative_2022} & 46$^e$ & 57$^e$ & 71$^e$ & 60$^e$ \\ 
RF &  Paper \cite{abdalgawad_generative_2022} & 40$^e$ & 54$^e$ & 62$^e$ & 57$^e$ \\ 
Hybrid (AAE + KNN) &  Paper \cite{abdalgawad_generative_2022} & 95$^e$ & 97$^e$ & 98$^e$ & 98$^e$ \\ 
Hybrid (BiGAN + KNN) &  Paper \cite{abdalgawad_generative_2022} & 95$^e$ & 98$^e$ & 97$^e$ & 97$^e$ \\ \hline

\hline
\textbf{Hybrid (RF, XGBoost and AdaBoost)} & \textbf{Ours} & \textbf{99} & \textbf{99} & \textbf{99} & \textbf{99}\\
\hline
\end{tabular}\\
\footnotesize{
$^a$, $^c$, and $^e$: Calculated by using the average of the values at the multi-classes provided in~\cite{prazeres_evaluation_2022},\cite{ullah_towards_2021}, and \cite{abdalgawad_generative_2022}, respectively.\\
$^b$: Calculated by using the average of the values at normal and attack cases provided in~\cite{elkashlan_machine_2023}.\\
$^d$: Calculated by using the average of the values at DDoS, C\&C, and PortScan attacks provided in~\cite{park_enhanced_2023}.\\
}
\end{table*}


\section{Conclusion}\label{sec:Section6}




The rapid expansion of the Internet of Things (IoT) has increased its vulnerability to sophisticated cyber threats, necessitating robust and scalable Intrusion Detection Systems (IDS).  This study introduces a hybrid approach to mitigate these threats, leveraging the IoT-23 dataset.  Our hybrid classifier models integrate tree-based algorithms—Random Forest (RF), XGBoost, and AdaBoost for multi-class classification, and RF, XGBoost, and K-Nearest Neighbors (KNN) for binary classification.  These models achieved exceptional performance, reaching 99.99\% accuracy, precision, recall, and F1-score across all metrics for binary classification, and 99\% for each metric in multi-class classification. This performance surpasses existing binary classification results and improves upon some multi-class metrics (as detailed in Tables I and II).  The scalability of our approach stems from the use of tree-based machine learning models in conjunction with a large dataset, enabling it to effectively handle the complexities of IoT environments.  The superior performance of our hybrid models compared to those in Table~\ref{Table.1} and Table~\ref{Table.2} demonstrates their suitability for IoT IDS.  

Future research will focus on real-time implementation, further feature selection optimization, adaptability to emerging cyber threats, and the integration of explainable AI to enhance transparency and trust in cybersecurity decision-making.

\balance
\bibliographystyle{IEEEtran}
\bibliography{references.bib}

\end{document}